\documentclass[aps, prd, showpacs, twocolumn, nofootinbib, preprintnumbers, superscriptaddress, groupedaddress]{revtex4}
\usepackage[utf8]{inputenc}
\usepackage{bm}
\usepackage{latexsym,amssymb,amsmath,amsfonts}
\usepackage{graphicx}
\usepackage{longtable}

\graphicspath{{fig/}}
\usepackage[usenames]{color}
\usepackage[breaklinks, backref, colorlinks=true]{hyperref}

\def \tsid {t_\mathrm{s}}
\def \iday {{i_\mathrm{day}}}
\def \sday {{T_\mathrm{s}}}

\begin{document}

 \preprint{LIGO-P1500007}
 
\title{Fast Gravitational Wave Radiometry using Data Folding}

\author{Anirban~Ain}
\email{ainz@iucaa.ernet.in}
\affiliation{Inter-University Centre for Astronomy and Astrophysics (IUCAA), Pune 411007, India}

\author{Prathamesh Dalvi}
\email{pratssd@gmail.com}
\affiliation{Birla Institute of Technology \& Science, Pilani - Goa Campus, India}

\author{Sanjit~Mitra}
\email{sanjit@iucaa.ernet.in}
\affiliation{Inter-University Centre for Astronomy and Astrophysics (IUCAA), Pune 411007, India}

\begin{abstract}

Gravitational Waves (GWs) from the early universe and unresolved astrophysical sources are expected to create a stochastic GW background (SGWB). The GW radiometer algorithm is well suited to probe such a background using data from ground based laser interferometric detectors. Radiometer analysis can be performed in different bases, e.g., isotropic, pixel or spherical harmonic. Each of these analyses possesses a common temporal symmetry which we exploit here to fold the whole dataset for every detector pair, typically a few hundred to a thousand days of data, to only one sidereal day, without any compromise in precision. We develop the algebra and a software pipeline needed to fold data, accounting for the effect of overlapping windows and non-stationary noise. We implement this on LIGO's fifth science run data and validate it by performing a standard anisotropic SGWB search on both folded and unfolded data.
Folded data not only leads to orders of magnitude reduction in computation cost, but it results in a conveniently small data volume of few gigabytes, making it possible to perform an actual analysis on a personal computer, as well as easy movement of data. A few important analyses, yet unaccomplished due to computational limitations, will now become feasible.
Folded data, being independent of the radiometer basis, will also be useful in reducing processing redundancies in multiple searches and provide a common ground for mutual consistency checks.
Most importantly, folded data will allow vast amount of experimentation with existing searches and provide substantial help in developing new strategies to find unknown sources.

\end{abstract}
  
\maketitle


\section{Introduction}
\label{intro}

A large number of sources in the universe are expected to emit short and long duration Gravitational Waves (GWs)~\cite{gw,MTW,ThorneG300}. Though we have convincing evidence of the existence of GWs~\cite{HulseTaylor,WeisbergNiceTaylor}, they have not yet been detected directly. A number of worldwide efforts are ongoing to detect GW signals of different kinds and at different wavelengths~\cite{AdvLIGO,AdvVirgo,KAGRA,IPTA,DECIGO,BBO}. The advanced ground based laser interferometric detectors, AdvLIGO~\cite{AdvLIGO}, AdvVIRGO~\cite{AdvVirgo} and KAGRA~\cite{KAGRA}, are the most promising candidates for the first detection of individual GW events, e.~g., signal from the coalescence of a compact binary. The detectors, however, will be sensitive to individual sources only from the nearby universe, perhaps up to few hundred Mpc to few Gpc~\cite{LIGOEventRate}, depending on the nature and strength of the sources. The rest of the sources in the universe, namely the faint and the distant ones, are likely to create a stochastic gravitational wave background (SGWB)~\cite{TanVuk08,CowTan06}. Early universe processes like inflation and phase transitions are also expected to contribute to the background~\cite{AllenSchool,Grishchuk00,Turner96}. If the background is dominated by the high redshift universe, it will likely be more isotropic. However, the collections of faint sources in the local universe, e. g., the huge population of milli-second pulsars in nearby galaxy clusters~\cite{hotspot}, may create a stronger astrophysical foreground~\cite{Mazumder2014} which will then make the net background predominantly anisotropic.

Detection of a SGWB is one of the most important focus areas of current observational astronomy~\cite{CMBTFR}. It would not only provide a ``smoking gun" test of inflation, but it can also provide information on average properties of various astrophysical objects not accessible to conventional electro-magnetic astronomy. Several upper limits have been placed on SGWB at different frequencies using data from a wide class of experiments, e.g., LIGO, Virgo, ALLEGRO, EPTA, WMAP, BICEP2, Planck~\cite{BBNBound,isoLVC_2009-10,sgwbS5iso,sgwbS5dir,ALLEGRO_limit,EPTA_Limit,SmithEtAl,RottiSouradeep,WMAP9par,PlanckParam,BICEP2,BICEP2-Planck,PlanckPolDust}.

Over the past three decades a robust algorithm to search for long duration SGWB, by cross-correlating data from pairs of ground-based GW detectors, has been developed~\cite{Michelson87, christ92, flan93, allen97, allen01, LazzariniWeiss, ballmer06, Mitra07, Thrane09} and implemented on real data~\cite{isoLVC_2009-10,sgwbS5iso,sgwbS4dir,sgwbS5dir,ALLEGRO_limit}. Though these searches are performed for isotropic and anisotropic backgrounds in different bases, they can be cast as a single unified maximum likelihood (ML) based radiometer analysis. The algebra then clearly reveals a temporal symmetry that could be utilised to fold data to one sidereal day. Data in this context are complex time-frequency maps incorporating outputs of pairs of detectors, which are often available as intermediate products of different SGWB searches. We can fold the time-frequency maps along the time axis modulo the sidereal day. Though in principle it is a simple scheme, the process becomes somewhat involved in order to account for overlapping Hanning windows used in data preprocessing to prevent spectral leakage~\cite{LazzariniRomano}.

We developed a parallel pipeline to implement folding on LIGO data. To test the method and the implementation we apply the pipeline to LIGO's fifth science run (S5) data. We validate the folded data by running a standard anisotropic SGWB search pipeline on it with nominal approximations and comparing the skymaps with the same obtained from unfolded data.

This paper is organised as follows. We begin with a brief review of the radiometer algorithm for a network of detectors in section~\ref{review}. The methodology for folding is described in section~\ref{folding}. Implementation and numerical results are presented in section~\ref{results}. We conclude in section~\ref{concl} with discussions on the method, the results and a list of important advantages that folded data can provide.


\section{GW Radiometer}
\label{review}

The GW radiometer algorithm~\cite{Michelson87, christ92, flan93, allen97, allen01, LazzariniWeiss, ballmer06, Mitra07, Thrane09} optimally probes either an isotropic or anisotropic SGWB and generates an all-sky map. The algorithm uses earth rotation aperture synthesis imaging, as is often used in radio astronomy. The same GW signal arrives at different times at geographically separated detectors. If the strain data from two detectors are cross-correlated, accounting for the delay between the sites in receiving the signals from a given direction, the signals add coherently whereas the noise does not. Thus, upon integration over a reasonably long observation time, the signal cross-correlation grows faster than noise variance, making the detection statistic more and more significant. The rotation of the earth provides different orientations of the detectors and the baselines, which makes it possible to make a map of the sky even with only two detectors.
%

The radiometer algorithm emerges naturally as the maximum likelihood solution to the probability of finding an anisotropy in the data from two detectors. A brief review of the algebra is given below. Based on this, it is straightforward to show how data can be folded for a radiometer search.

Let us consider a two detector ``baseline'' $I$ constituted by the detectors denoted by $\mathcal{I}_1$ and $ \mathcal{I}_2$ (calligraphic letter denotes detectors in the corresponding baseline). Time-series data $s_{\mathcal{I}_{1, 2}}(t)$ from the detectors are the sum of the signal $h_{\mathcal{I}_{1, 2}}(t)$ and the noise $n_{\mathcal{I}_{1, 2}}(t)$,
\begin{eqnarray}
s_{\mathcal{I}_{1}}(t) \ = \ h_{\mathcal{I}_{1}}(t) \ + \ n_{\mathcal{I}_{1}}(t)\label{det_output1} \, ,\\
s_{\mathcal{I}_{2}}(t) \ = \ h_{\mathcal{I}_{2}}(t) \ + \ n_{\mathcal{I}_{2}}(t)\label{det_output2} \, .
\end{eqnarray}
It is more convenient to divide the data into smaller time segments and to work with their `Short-term Fourier Transforms' (SFTs). The SFT of a time series $s(t)$ for a segment of duration $\tau$ is given by
\begin{equation}
\tilde {s}(t;f) \ := \ \int_{t-\tau/2}^{t+\tau/2} \mathrm{d}t' \, s(t') \, e^{-i2\pi f t'} \, .
\label{SFT}
\end{equation}
The argument $t$ in $\tilde {s}(t;f)$ is a timestamp to mark a specific segment. Unless otherwise specified, we represent the frequency domain Fourier Transform of a time series by putting a $\tilde{\ }$ on top of the corresponding variable.

Statistically the noise in a detector is expected to be uncorrelated with the GW strain in that detector, or the noise in another geographically well separated detector.\footnote{There can be small traces of correlated noise present in two geographically separated detectors caused by global magnetic fields from Schumann resonances~\cite{magNoise}. Methods have been proposed to mitigate those noises~\cite{noiseCorrMitigation}. The same prescription can be followed for generation and use of folded data.}
Hence, one has
\begin{eqnarray}
\langle \tilde{n}^*_{\mathcal{I}_{1,2}}(t;f) \, \tilde{h}_{\mathcal{I}_{1,2}}(t;f) \rangle \ = \ 0 \, ,\\
\langle \tilde{n}^*_{\mathcal{I}_{1}}(t;f) \, \tilde{n}_{\mathcal{I}_{2}}(t;f) \rangle \ = \ 0 \, .
\label{eq:noiseUnorr}
\end{eqnarray}

Since a stochastic background is characterised by the second moment of the signal (power spectral density) and cross-correlation techniques are used to probe a signal, we start our analysis with the cross-power spectral density (CSD) of data from a baseline. CSD is defined as the product of the complex conjugate of SFT of one detector's data for a certain segment with the SFT of data from the other detector for the same segment. Thus, for a baseline $I$, the CSD ($\mathbf{C}^I$) and the corresponding noise ($\mathbf{n}^I$) in the small signal limit, $\langle |\tilde{h}(t;f)|^2 \rangle \ll \langle |\tilde{n}(t;f)|^2 \rangle$, are given by
\begin{eqnarray}
\mathbf{C}^I \ \equiv \ C^I_{ft} &:=& \widetilde{s}_{\mathcal{I}_1}^*(t;f) \, \widetilde{s}_{\mathcal{I}_2}(t;f)\label{eq:CSD} \, ,\\
\mathbf{n}^I \ \equiv \ n^I_{ft} &:=& \widetilde{n}_{\mathcal{I}_1}^*(t;f) \, \widetilde{n}_{\mathcal{I}_2}(t;f)\label{eq:noise} \, .
\end{eqnarray}
The expected instantaneous cross-power noise variance is given by
\begin{equation}
\sigma_{Ift}^{2} \ := \ \langle n^{I*}_{ft} \, n^I_{ft} \rangle \ = \ \frac{\tau^2}{4} P_{\mathcal{I}_1}(t;f) \, P_{\mathcal{I}_2}(t;f) \, ,
\label{eq:varCSD}
\end{equation}
where $P_{\mathcal{I}_{1,2}}(t;f)$ are the one-sided power spectral density (PSD) of noise $n_{\mathcal{I}_{1,2}}$ for segment $t$ and $\tau$ is the duration of a segment.

An SGWB can be modeled by the expected shape of its frequency PSD $H(f)$ and its expected amplitude $\mathcal{P}(\mathbf{\widehat{\Omega}})$ in the direction of the sky $\mathbf{\widehat{\Omega}}$.~\footnote{Note that, so far SGWB search algorithms have been developed only for cases where the shape of the frequency spectrum $H(f)$ is constant in every direction, only the amplitude $\mathcal{P}(\mathbf{\widehat{\Omega}})$ varies. In a more general scenario the shape of the power spectrum can also depend on direction $\mathbf{\widehat{\Omega}}$. In principle, a targeted radiometer search can be applied in such cases, given a model for $H(f)$ for every direction, though it has not been tested yet.}
One can expand the ``SGWB skymap'' $\mathcal{P}(\mathbf{\widehat{\Omega}})$ in any chosen basis $e_\alpha(\mathbf{\widehat{\Omega}})$, such that,
\begin{equation}
\mathcal{P}(\mathbf{\widehat{\Omega}}) \ := \ \sum_\alpha \mathcal{P}_\alpha \, e_\alpha(\mathbf{\widehat{\Omega}}) \, .
\end{equation}
So far in literature the following bases have been used for SGWB searches: $e_\alpha(\mathbf{\widehat{\Omega}}) = 1$ (isotropic search, $\alpha = 0$), $e_\alpha(\mathbf{\widehat{\Omega}}) = \delta(\Omega-\Omega_\alpha)$ (pixel basis, $\alpha$ is the pixel index) and $e_\alpha(\mathbf{\widehat{\Omega}}) = Y_{lm}$ (spherical harmonic basis, $\alpha \equiv lm$).
In general, in any basis, including the ones mentioned above, the expectation of the CSD from the baseline $I$ can be written as~\cite{Thrane09}
\begin{equation}
\langle C^I_{ft} \rangle \ := \ \tau \, H(f) \, \sum_\alpha  \mathcal{P}_\alpha \, \gamma^{I}_{ft,\alpha} \, ,
\label{eq:meanCSD}
\end{equation}
where the general overlap reduction function
\begin{equation}
\label{ov_rd_fn}
\gamma_{ft,\alpha}^{I} := \sum_{A} \int_{S^2} d \mathbf{\hat \Omega} F^{A}_{\mathcal{I}_1}(\mathbf{\hat \Omega},t) F^{A}_{\mathcal{I}_2}(\mathbf{\hat\Omega},t) e^{2\pi i f\frac{\mathbf{\hat \Omega}\cdot {\mathbf{\Delta x}_I (t)}}{c}} e_{\alpha}(\mathbf{\hat \Omega}) \, ,
\end{equation}
and $A$ is the polarisation ($+$ or $\times$). $\gamma_{ft,\alpha}^{I} $ contains all the information about the antenna pattern functions $F^{A}_{\mathcal{I}_{1,2}}(\mathbf{\hat \Omega},t)$ of the detectors, baseline separation ${\mathbf{\Delta x}_I (t)}$ and the basis $e_{\alpha}(\mathbf{\hat \Omega})$. It is worth noting that for small enough segment duration ($\tau \sim 1$ minute), $\gamma_{ft,\alpha}^{I}$ varies insignificantly with $t$ within a segment~\cite{sgwbS4dir}.


%
From the above discussions, combining Eqs.~(\ref{eq:CSD}), (\ref{eq:noise}) \& (\ref{eq:meanCSD}), one can express the observed CSD $\mathbf{C}^{I}$ from a given baseline $I$ as a linear convolution of the true sky $\bm{\mathcal{P}} \equiv \mathcal{P}_\alpha$,
\begin{equation}
\mathbf{C}^{I} \ = \ \mathbf{K}^{I} \cdot \bm{\mathcal{P}} \ + \ \mathbf{n}^{I} \, ,
\end{equation}
through a deterministic kernel (a.~k.~a. the beam function in pixel basis) [see Eq.~(\ref{eq:meanCSD})]
\begin{equation}
\mathbf{K}^I \ \equiv \ K^{I}_{ft,\alpha} \ := \ \tau \, H(f) \, \gamma^{I}_{ft,\alpha} \, ,
\end{equation}
and additive noise $\mathbf{n^{I}} \equiv n^{I}_{ft}$ with covariance (in the small signal limit) [see Eq.~(\ref{eq:varCSD})]
\begin{equation}
\bm{\sigma}^I  \equiv  \sigma^{I}_{ft, f't'}  \ \approx \ \mathrm{Cov}(n^I_{ft}, n^{I}_{f't'})  =  \delta_{tt'} \delta_{ff'} \sigma_{Ift}^{2} \, .
\end{equation}
For the case of multiple (say $N_b$) baselines, one can combine the data, noise and the kernel as independent observations respectively as
\begin{equation}
\mathbf{C} \ := \ \left( \begin{array}{l}
\mathbf{C}^{1} \\ \mathbf{C}^{2} \\ \vdots \\ \mathbf{C}^{N_{\rm b}}
\end{array} \right);\ \
\mathbf{n} \ := \ \left( \begin{array}{l}
\mathbf{n}^{1} \\ \mathbf{n}^{2} \\ \vdots \\ \mathbf{n}^{N_{\rm b}}
\end{array} \right);\ \
\mathbf{K} \ := \ \left[ \begin{array}{l}
\mathbf{K}^{1} \\ \mathbf{K}^{2} \\ \vdots \\ \mathbf{K}^{N_{\rm b}}
\end{array} \right] \, . \nonumber
\end{equation}
The noise covariance matrix still remains block diagonal,
\begin{equation}
\begin{split}
\mathbf{N} \equiv N_{Ift,I'f't'} &= \langle \widetilde{n}_{\mathcal{I}_1}(t;f) \widetilde{n}^*_{\mathcal{I}_2}(t;f) \widetilde{n}_{\mathcal{I}'_1}^*(t';f') \widetilde{n}_{\mathcal{I}'_2}(t';f') \rangle\\
&= \delta_{II'} \delta_{tt'} \delta_{ff'} \sigma_{Ift}^{2} \, ,
\end{split}
\end{equation}
as the expectation vanishes if either $\mathcal{I}_1 \ne \mathcal{I}'_1$ or $\mathcal{I}'_2 \ne \mathcal{I}_2$.
%
%
Combining the data vectors in this manner retains the single baseline form of the convolution equation,
\begin{equation}
\mathbf{C} \ = \ \mathbf{K} \cdot \bm{\mathcal{P}} \ + \ \mathbf{n} \, .
\end{equation}

The objective of a search is to estimate the coefficients $\mathcal{P}_\alpha$ from the CSDs from one or more baselines.
A standard ML solution to the above convolution equation, which provides an estimate of the true sky~\cite{Mitra07}, can be written as
\begin{equation}
\mathcal{\hat{P}}_\alpha \ \equiv \ \hat{\bm{\mathcal{P}}} \ = \ \mathbf{\Gamma}^{-1} \cdot \mathbf{X} \, ,
\end{equation}
where $\mathbf{X} \equiv X_\alpha$ is called the dirty map,
\begin{equation}
\label{eq:unfX}
\begin{split}
\mathbf{X} :=& \mathbf{K}^\dagger \cdot \mathbf{N}^{-1} \cdot \mathbf{C} \quad \Rightarrow \ X_\alpha = \sum_{Ift} K^{I*}_{ft,\alpha} \sigma_{Ift}^{-2} C^I_{ft} \\
= \ & \frac{4}{\tau} \sum_{Ift} \frac{ H(f) \gamma^{I*}_{ft,\alpha}} {P_{\mathcal{I}_1}(t;f) P_{\mathcal{I}_2}(t;f)} \widetilde{s}_{\mathcal{I}_1}^*(t;f) \widetilde{s}_{\mathcal{I}_2}(t;f) \, ,
\end{split}
\end{equation}
and $\mathbf{\Gamma} \equiv \Gamma_{\alpha\alpha'}$ is the Fisher information matrix,
\begin{equation}
\label{eq:unfGamma}
\begin{split}
\mathbf{\Gamma} :=& \mathbf{K}^\dagger \cdot \mathbf{N}^{-1} \cdot \mathbf{K} \quad \Rightarrow  \ \Gamma_{\alpha\alpha'} =\ \sum_{Ift} K^{I*}_{ft,\alpha}\, \sigma_{Ift}^{-2}  \, K^{I}_{ft,\alpha'} \\ 
= \ & 4 \sum_{Ift} \frac{H^2(f)}{P_{\mathcal{I}_1}(t;f) \, P_{\mathcal{I}_2}(t;f)} \,\gamma^{I*}_{ft,\alpha} \, \gamma^{I}_{ft,\alpha'} \, .
\end{split}
\end{equation}
Thus computing $\mathbf{X}$ and $\mathbf{\Gamma}$ from data in a chosen basis is the essential goal of an SGWB search.


\section{Folding Data to One Sidereal Day}
\label{folding}

\subsection{General Formalism}

The idea behind folding follows trivially from the formulae for the ML estimation presented in the previous section. Since we are applying earth rotation synthesis imaging, the kernel $K^{I}_{ft,\alpha}$ has a period of one sidereal day (i.e. 23 hr 56 min 4 sec). This symmetry can be exploited to fold the non-periodic part, the entire data of several hundreds of days, to one sidereal day.

\begin{figure}[h]
\includegraphics[width=0.49\textwidth]{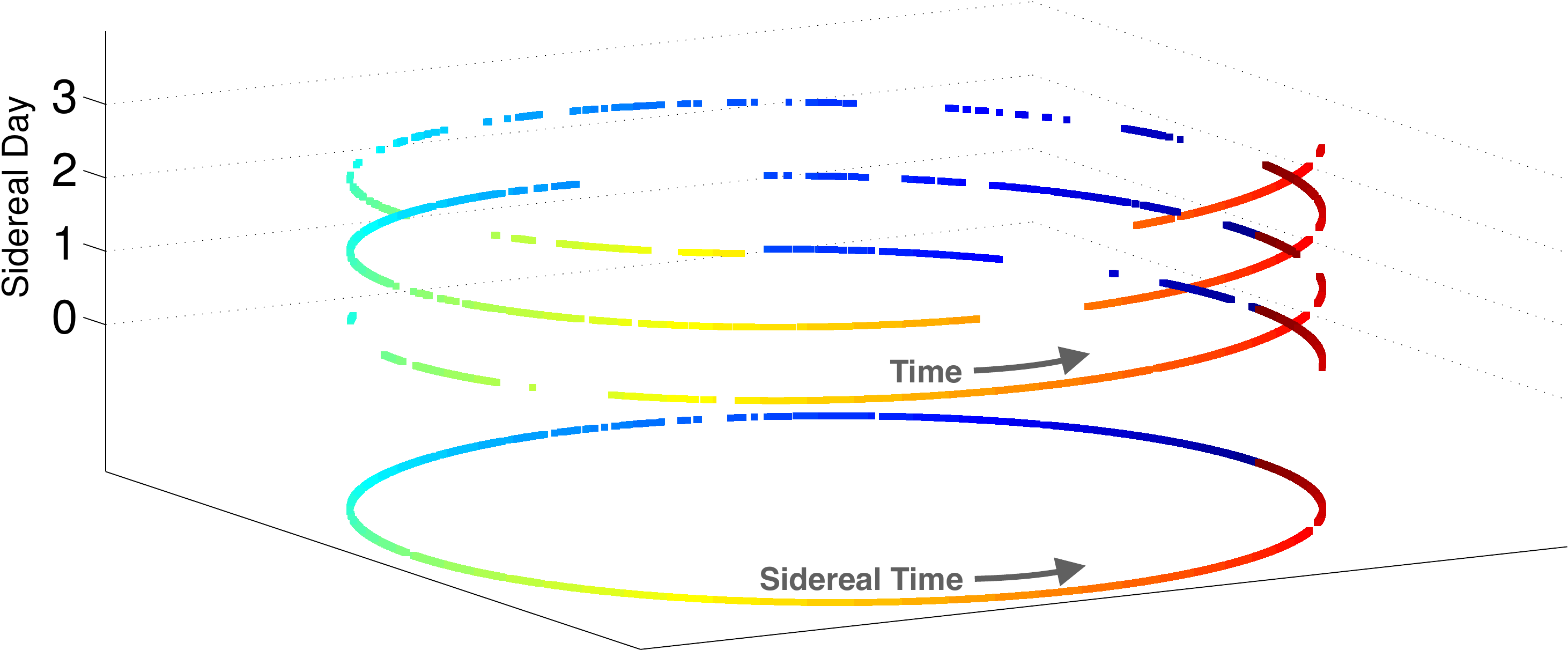}
\caption{\label{fig:spiral}This diagram illustrates the folding scheme using three days of LIGO S5 data from GPS time $860832366$~sec to $861090858$~sec. Each point on the spiral represents one segment marked by a GPS time and its color represents the corresponding sidereal time. The projected ring at the bottom represents folded data. Missing points in the spiral and the projected ring represent missing data. Had we considered many more days of data in this plot, the gaps in the projected ring would have disappeared. As we will see, folding generates a ring for each frequency bin, that is, a time-frequency map.}
\end{figure}

Both the quantities needed for ML estimation of $\mathcal{P}_\alpha$, the dirty map $X_\alpha$ and the Fisher information matrix $\Gamma_{\alpha\alpha'}$, involve summations over time segments marked by $t$. These summations can be split into two parts using $t = \iday \times \sday + \tsid$, where $\iday$ is a dimensionless integer representing the sidereal day number in which a given $t$ lies, $\tsid$ is the remainder within that sidereal day and $\sday$ is the duration of one sidereal day.
Equivalently, the summation over $t$ in Eq.~(\ref{eq:unfX}) and Eq.~(\ref{eq:unfGamma}) can be broken into two summations, $\Sigma_t \rightarrow \Sigma_{\iday} \Sigma_{\tsid}$, where $\iday$ runs over of the total number of sidereal days for which data is being processed and $\tsid$ runs over one sidereal day. This is schematically presented in Figure~\ref{fig:spiral}.

Following the above convention, it is straightforward to introduce folding by rewriting Eq.~(\ref{eq:unfX}) and Eq.~(\ref{eq:unfGamma}) as
%
%
\begin{eqnarray}
X_\alpha & = & \sum_{If\tsid} K^{I*}_{f\tsid,\alpha} \,  x^I_{f\tsid} \, , \label{eq:fX}\\
\Gamma_{\alpha\alpha'}  & = & \sum_{If\tsid} K^{I*}_{f\tsid,\alpha} \, K^{I}_{f \tsid,\alpha'} \, v^I_{f\tsid} \, , \label{eq:fGamma}
\end{eqnarray}
where the quantities summed over all sidereal days,
\begin{eqnarray}
x^I_{f\tsid} & = & \sum_{\iday} \sigma_{If(\iday\sday + \tsid)}^{-2} \,C^{I}_{f(\iday\sday + \tsid)} \, , \label{eq:fx} \\
v^I_{f\tsid} & = & \sum_{\iday} \sigma_{If(\iday\sday + \tsid)}^{-2} \, , \label{eq:fv}
\end{eqnarray}
are the {\em folded data}, which span only one sidereal day. Thus, without any loss of generality or precision in deriving the final results $X_\alpha$ and $\Gamma_{\alpha\alpha'}$, the long time-frequency maps $C^I_{ft}$ (complex) and its variance $\sigma^2_{Ift}$ (real) for each baseline are now compressed into one-sidereal-day-long time-frequency maps $x^I_{f\tsid}$ (complex) and $v^I_{f\tsid}$ (real). Note that non-stationarity of noise has automatically been taken into account in the above formulae through the inverse variance weights, $\sigma_{If(\iday\sday + \tsid)}^{-2}$, in Eq.~(\ref{eq:fv}).

Each segment of folded and unfolded data has the same number of complex and real vectors, hence data compression is exactly given by the ratio of the number of segments in unfolded and folded data. Computational speed up is also given by the same ratio, as it is roughly proportional to the data volume in a  radiometer analysis.

Since Eq.~(\ref{eq:fx}) and Eq.~(\ref{eq:fv}) do not involve any quantity with an index $\alpha$, {\em folded data is independent of the search basis}. Hence, similar to the unfolded time-frequency maps, it is possible to perform different types of searches on folded data by making different choices of the SGWB spectrum $H(f)$ and overlap reduction function $\gamma^{I}_{ft,\alpha}$.
 
\subsection{Correction for Overlapping Window}

A smooth window function is often applied to the time series data from the detectors to reduce leakage from strong spectral lines in the SFTs. Applying such a window naively, would, however, lead to effective loss of data, as major portion of quality data does not receive full weighting. To prevent this, the windows are made to overlap. A method was developed for using $50$\% overlapping Hanning windows in SGWB analyses~\cite{LazzariniRomano} and was applied to LIGO-Virgo stochastic searches~\cite{isoLVC_2009-10,sgwbS5iso,sgwbS5dir}. Accounting for overlapping windows makes the algebra for folding considerably involved, though the end result is fairly simple and elegant, which we present below.

Overlapping windows introduce a correlation between SFTs from neighbouring segments, hence the noise covariance matrix becomes tridiagonal,
\begin{equation}
\label{eq:noiseCov}
\begin{split}
 N_{Ift, I'f't'} \ & = \ \delta_{II'} \, \delta_{ff'} [\delta_{tt'} \, \sigma_{Ift}^2 \, + \\
 & \, \delta_{(t-1)t'} \, \sigma^{I}_{ft, f(t-1)} \, + \, \delta_{(t+1)t'} \, \sigma^{I}_{ft, f(t+1)}] \, .
 \end{split}
\end{equation}
Note that, for brevity of notations $t$ has been used here as an integer index to time segments, hence $t \pm 1$ represent respectively the next and the previous segments. To convert $t$ to proper units of time, one needs to multiply it by the segment duration $\tau$.
The non-zero off-diagonal components of the covariance of CSD, $\bm{\sigma}^I \equiv \sigma^I_{ft,f't'}$, for $50$\% overlap and nearly white noise are given by~\cite{LazzariniRomano}
\begin{eqnarray}
 &&\sigma^{I}_{ft, f(t\pm 1)} \ =  \ \varepsilon^I_{t\pm 1} \, [\sigma^2_{Ift} \ + \ \sigma^2_{If(t\pm 1)}]/2, \label{eq:sigmaft} \\
 && \varepsilon^I_{t\pm 1} \ = \ \left\{ \begin{array}{ll}
 \mathcal{W}_I & \mbox{if segment $t\pm 1$ exists for baseline $I$}\\
 0 & \mbox{otherwise}
 \end{array} \right. ,\nonumber
\end{eqnarray}
and the coefficient $\mathcal{W}_I$ depends on the functional form of the window function. If $w_{\mathcal{I}_1}(t)$ and $w_{\mathcal{I}_2}(t)$ are the window functions applied to the time series data from the respective detectors, the coefficient is given by
\begin{equation}
\mathcal{W}_I \ = \ \frac{\overline{w_{\mathcal{I}_1}(t) \, w_{\mathcal{I}_1}(t+\tau/2) \, w_{\mathcal{I}_2}(t) \, w_{\mathcal{I}_2}(t+\tau/2)}}{2 \, \overline{w_{\mathcal{I}_1}^2(t) \, w_{\mathcal{I}_2}^2(t)}} \, ,
\end{equation}
where the bar on the quantities denotes average over time. Since the window function remains the same for the whole stretch of the data, the averaging needs to be done over a single segment only. For the same Hanning window on both the data streams, this factor turns out to be $\mathcal{W}_I \approx 3/70$.

To proceed further we require the inverse of the covariance matrix $\mathbf{N}$. A tridiagonal matrix of the form
\begin{equation}
\mathbf{A} \ \equiv \ A_{ij} \ = \ \alpha_i \, \delta_{ij} \ + \ \beta^+_i \delta_{i(j+1)} \ + \ \beta^-_i \delta_{i(j-1)} \, ,
\end{equation}
such that $|\beta^\pm_i/\alpha_j| \ll 1$, for all $i,j$, when multiplied to the matrix
\begin{equation}
\mathbf{B} \ \equiv \ B_{jk} \ = \ \frac{1}{\alpha_j} \left[ \delta_{jk} \ - \ \frac{\beta_{j}^+}{\alpha_k} \delta_{j(k+1)} \ - \ \frac{\beta_j^-}{\alpha_k} \delta_{j(k-1)} \right] \, ,
\end{equation}
one gets,
\begin{equation}
\sum_{j} A_{ij} \, B_{jk} \ = \ \delta_{ik} \ + \ \mathcal{O}(|\beta^\pm/\alpha|^2) \, .
\end{equation}
Hence, matrix $\mathbf{B}$ can be regarded as the inverse of matrix $\mathbf{A}$ to first order in $|\beta_\pm/\alpha|$. We use this to invert the matrix $\bm{\sigma}^I$ to first order in $\mathcal{W}_I$ (accuracy $\sim \mathcal{W}_I^2 \approx 0.2\%$). By substituting $\alpha_t = \sigma^{2}_{Ift}$, $\beta^\pm_t = \sigma^{I}_{ft, f(t\pm 1)}$ and Eq.~(\ref{eq:sigmaft}), one gets,
\begin{equation}
\begin{split}
[\sigma^I]^{-1}_{ft,ft} \ = &\ 1/\alpha_t \ = \ \sigma^{-2}_{Ift} \, \\
{[\sigma^I]}^{-1}_{ft, f(t\pm 1)} \ \approx&\  - \frac{\beta^\pm_t}{\alpha_t \alpha_{t\pm 1}} \ = \ - \frac{\varepsilon^I_{t\pm 1}}{2} \, [\sigma_{Ift}^{-2} \ + \  \sigma_{If(t\pm 1)}^{-2}] \, ,
\end{split}
\end{equation}
all other elements of $[\bm{\sigma}^I]^{-1}$ are zero. In turn, the inverse of the (block diagonal) matrix $\mathbf{N}$ [Eq.~(\ref{eq:noiseCov})] becomes,
\begin{equation}
\begin{split}
[\mathbf{N}]^{-1} \ &\equiv \ [N]^{-1}_{Ift, I'f't'} \approx \ \delta_{II'} \, \delta_{ff'} \big[ \delta_{tt'} \, [\sigma^{I}]^{-1}_{ft,ft} \, + \ \\ & \delta_{(t-1)t'} \, [\sigma^{I}]^{-1}_{ft, f(t-1)} +  \delta_{(t+1)t'} \, [\sigma^{I}]^{-1}_{ft, f(t+1)} \big] \, .
\end{split}
\end{equation}

We now follow a convention often used in LIGO-Virgo analysis to avoid unnecessary factors arising from the Hanning windows in the expectation of the statistic. We redefine the CSD as,
\begin{equation}
\mathbf{C}^I \ \equiv \ C^I_{ft} \ := \ \frac{1}{\overline{w_{\mathcal{I}_1}(t) w_{\mathcal{I}_2}(t)}}  \widetilde{s}_{\mathcal{I}_1}^*(t;f) \, \widetilde{s}_{\mathcal{I}_2}(t;f) \, ,
\end{equation}
where $\widetilde{s}_{\mathcal{I}_{1,2}}(t;f)$ are windowed SFTs. The variance then becomes
\begin{equation}
\sigma^2_{Ift}  \ = \  \frac{\overline{w_{\mathcal{I}_1}^2(t) w_{\mathcal{I}_2}^2(t)}}{\left[\overline{w_{\mathcal{I}_1}(t) w_{\mathcal{I}_2}(t)}\right]^2} \, \frac{\tau^2}{4} \, P_{\mathcal{I}_1}(t;f) \, P_{\mathcal{I}_2}(t;f) \, .
\end{equation}
All the other formulae above remain unchanged, as the quantities appearing in the final formulae have been expressed in terms of $C^I_{ft}$ and $\sigma^2_{Ift}$.

\begin{widetext}
Using these results, the expressions for dirty map $\mathbf{X}$ and Fisher information matrix $\mathbf{\Gamma}$ [Eqs.~(\ref{eq:unfX}) and (\ref{eq:unfGamma})] become,
\begin{eqnarray}
X_\alpha &=& \sum_{Ift} K^{I*}_{ft,\alpha} \left[  \sigma^{-2}_{Ift} \, C^{I}_{ft} \ - \  \frac{1}{2} \varepsilon^I_{t-1} \, \left\{ \sigma_{Ift}^{-2} \, + \, \sigma_{If(t-1)}^{-2} \right\} \, C^{I}_{f(t-1)} \right. 
 - \ \left. \frac{1}{2} \varepsilon^I_{t+1} \, \left\{\sigma_{Ift}^{-2} \, + \, \sigma_{If(t+1)}^{-2} \right\} \, C^{I}_{f(t+1)}\right] \, ,\\
 \Gamma_{\alpha\alpha'} &=& \sum_{Ift} K^{I*}_{ft,\alpha}\left[\sigma^{-2}_{Ift} \, K^I_{ft,\alpha'} \, - \,  \frac{1}{2}\right. \varepsilon^I_{t-1} \, \left\{\sigma_{Ift}^{-2} \, + \, \sigma_{If(t-1)}^{-2} \right\} \, K^I_{f(t-1),\alpha'} \, - \, \frac{1}{2} \varepsilon^I_{t+1}\left\{\sigma_{Ift}^{-2} \right.
 \left. \left.  + \, \sigma_{If(t+1)}^{-2} \right\} \, K^I_{f(t+1),\alpha'} \right] . \quad \
\end{eqnarray}
\end{widetext}

The above formulae apply to the standard (unfolded) radiometer analyses. We now write them in terms of folded data. We again partition the summation over $t$ to summations over $\iday$ and $\tsid$. The formulae for dirty map and Fisher information matrix, Eqs.~(\ref{eq:fX}) \& (\ref{eq:fGamma}), can still be cast in a way that they take simple forms,
\begin{eqnarray}
X_\alpha &=& \sum_{If\tsid} K^{I*}_{f\tsid,\alpha} \,  x^I_{f\tsid} \, , \label{eq:fwX}\\
\Gamma_{\alpha\alpha'} &=& \sum_{If\tsid} K^{I*}_{f\tsid,\alpha} \big[ K^I_{f\tsid,\alpha'}  \, v^I_{f\tsid} \label{eq:fwGamma} \\
&& - \ K^I_{f(\tsid-1),\alpha'}  \, u^I_{f\tsid} \ - \ K^I_{f(\tsid+1),\alpha'}  \, w^I_{f\tsid} \big] \, , \nonumber
\end{eqnarray}
where the expressions for $x^I_{f\tsid}$ and $v^I_{f\tsid}$, provided in Eqs.~(\ref{eq:fx}) \& (\ref{eq:fv}) for the unwindowed case, have been modified with correction terms and two more folded sets, $u^I_{f\tsid}$ and $w^I_{f\tsid}$, have been introduced. The final expressions for the {\em three real} folded data are given by
\begin{equation}
\begin{split}
v^I_{f\tsid} \ = &\  \sum_\iday \sigma^{-2}_{If(\iday\sday + \tsid)} \, , \\
u^I_{f\tsid} \ = &\  \sum_\iday \frac{1}{2} \varepsilon^I_{\iday\sday + \tsid-1} \ \times \\
& \qquad \left[ \sigma_{If(\iday\sday + \tsid)}^{-2} \, + \, \sigma_{If(\iday\sday + \tsid-1)}^{-2} \right] \, , \\
w^I_{f\tsid} \ = &\  \sum_\iday \frac{1}{2} \varepsilon^I_{\iday\sday + \tsid+1} \ \times\\
& \qquad \left[ \sigma_{If(\iday\sday + \tsid)}^{-2} \, + \, \sigma_{If(\iday\sday + \tsid+1)}^{-2} \right] \, ,
\end{split}
\end{equation}
\begin{widetext}
and {\em one complex} folded time-frequency map
\begin{equation}
\begin{split}
x^I_{f\tsid} \ = &\ \sum_\iday \left[  \sigma^{-2}_{If(\iday\sday + \tsid)} \, C^{I}_{f(\iday\sday + \tsid)} - \ \frac{1}{2} \varepsilon^I_{\iday\sday + \tsid-1} \, \left\{ \sigma_{If(\iday\sday + \tsid)}^{-2} \, + \, \sigma_{If(\iday\sday + \tsid-1)}^{-2} \right\} \, C^{I}_{f(\iday\sday + \tsid-1)} \right. \\
&- \ \left. \frac{1}{2} \varepsilon^I_{\iday\sday + \tsid+1} \, \left\{\sigma_{If(\iday\sday + \tsid)}^{-2} \, + \, \sigma_{If(\iday\sday + \tsid+1)}^{-2} \right\} \, C^{I}_{f(\iday\sday + \tsid+1)} \right] \, .
\end{split}
\end{equation}
\end{widetext}

Although we made three real folded datasets ($u$, $v$, $w$) to ensure exact analysis, in practice, one set, a combination of them, can provide adequate accuracy. One can do this by replacing the kernels for the adjacent segments in Eq.~(\ref{eq:fwGamma}), $K^I_{f(\tsid \pm 1),\alpha}$, by the kernel for the central segment, $K^I_{f\tsid,\alpha}$. This is possible not only because the kernels in neighbouring segments differ by $\sim 10$\% at $f \sim 1$~kHz (lesser for smaller frequencies), provided the segment duration is $\sim 1$~min, so that the earth has not turned significantly during the combined interval of $\sim 2$~min spanned by three consecutive overlapping segments, but also the terms $K^I_{f(\tsid \pm 1),\alpha}$ are weighted by $\mathcal{W}_I$ in Eq.~(\ref{eq:fwGamma}), so the error introduced by this approximation is less than a percent. Then the expression for Fisher information matrix, Eq.~(\ref{eq:fwGamma}), simplifies to,
\begin{equation}
\Gamma_{\alpha\alpha'} \ \approx \ \sum_{If\tsid} K^{I*}_{f\tsid,\alpha} K^I_{f\tsid,\alpha'} \left[  v^I_{f\tsid}  - u^I_{f\tsid} - w^I_{f\tsid} \right] \, .
\label{eq:fwGammaApprox}
\end{equation}
Hence, instead of $u^I_{f\tsid}$, $v^I_{f\tsid}$ and $w^I_{f\tsid}$, only one quantity,
\begin{equation}
\label{eq:barv}
\bar{v}^I_{f\tsid} \ := \ v^I_{f\tsid}  \, - \, u^I_{f\tsid} \, - \, w^I_{f\tsid} \, ,
\end{equation}
suffices to provide adequate accuracy.
%
In addition, if noise can be taken as sufficiently stationary, such that the neighbouring segments have nearly the same PSDs, one could further simplify the folded data as,
\begin{equation}
\begin{split}
&\bar{v}^I_{f\tsid} = \sum_\iday \sigma^{-2}_{If(\iday\sday + \tsid)} (1-\varepsilon^I_{\iday\sday + \tsid - 1}-\varepsilon^I_{\iday\sday + \tsid +1}) \, ,\\
&x^I_{f\tsid} = \sum_\iday  \sigma^{-2}_{If(\iday\sday + \tsid)} \left[ C^{I}_{f(\iday\sday + \tsid)} \ - \ \right. \label{eq:approxfsid}\\
& \ \varepsilon^I_{\iday\sday + \tsid - 1} C^{I}_{f(\iday\sday + \tsid - 1)} - \left. \varepsilon^I_{\iday\sday + \tsid + 1} C^{I}_{f(\iday\sday + \tsid + 1)} \right] .
\end{split}
\end{equation}

These approximations cause insignificant ($< 1\%$) differences in the final results, yet simplify the analysis pipeline. Hence, it is a common practice by the LIGO community to include them in radiometer analyses. However, we emphasise that the folding method does not require these approximations. If necessary, one can develop a modified pipeline to make use of all the folded datasets for a more exact analysis.

It is also worth noting that the time-frequency maps $[\bar{v}^I_{f\tsid}]^{1/2}$ and $\bar{v}^I_{f\tsid} {x}^I_{f\tsid}$ can be used respectively as effective noise PSDs and CSD of two detectors. This representation of folded data can be readily incorporated into standard analysis pipelines, as done in section~\ref{results}.


\section{Implementation and Results}
\label{results}

We have developed a code for folding data to one sidereal day and implemented it on real data. To demonstrate the method and validate the results we apply a standard analysis code developed by the LIGO Scientific Collaboration to search for a stochastic background on both folded and regular unfolded data. For this purpose we have used LIGO's fifth science run data from the Hanford and the Livingston detectors.

Data from the detectors are stored in 128 sec long frame files. Raw data are down-sampled from 16384 Hz to 2048 Hz. Then a high pass filter is used to allow frequencies above 32 Hz to reduce seismic noise. The SFTs of the data from the two sites are then combined to obtain the CSDs, which are written to a new set of frame files with a reduced set of metadata. These new frames are called Stochastic Intermediate Data (SID). Each file contains one SID frame of duration $52$ seconds. This duration was chosen because one sidereal day is approximately $86164$ seconds long, which is an integral multiple of $52$. A frequency resolution of $\Delta f = 0.25$ was used. The data was windowed to avoid spectral leakage of strong instrumental lines with Hann function with 50\% overlap.

Our code folds the SIDs to Folded Stochastic Intermediate Data (FSID). The FSID frames are written in such a format that the standard LIGO analysis pipeline can be directly used to analyse this data. To match the pipeline, we make use of the slightly approximate folded data given by Eq.~(\ref{eq:approxfsid}) to create effective noise PSDs and CSD described above. Note that the FSID frames we generate also store the four folded datasets, $u$, $v$, $w$, $x$, needed to perform an exact analysis, if ever necessary, though it will require some changes in the existing pipeline.

We first run the pipeline on SID frames for a certain observation period to search for an anisotropic SGWB in the spherical harmonic (SpH) basis~\cite{sgwbS5dir}. The pipeline applies the same algebra that was presented in Section~\ref{review} and \citet{Thrane09}. We choose a source PSD $H(f)=(f/f_R)^\beta$, where the parameters $f_R$ and $\beta$ are respectively the reference frequency and spectral index for the (assumed) power-law behaviour of the GW spectrum. Although this validation study is independent of the parameters used in the analysis, we choose a set which is commonly used in LIGO-Virgo analyses~\cite{isoLVC_2009-10,sgwbS5iso,sgwbS5dir}. We search for a flat spectrum by setting $\beta=0$ and $f_R = 100$ Hz (redundant for $\beta=0$). We choose a spherical harmonic multipole cut-off of $l_\text{max} = 15$. The processed output comes in the form of a dirty map and the Fisher information matrix for individual jobs, which can be combined and further post-processed to make a skymap.

\begin{figure*}
\includegraphics[width=0.95\textwidth]{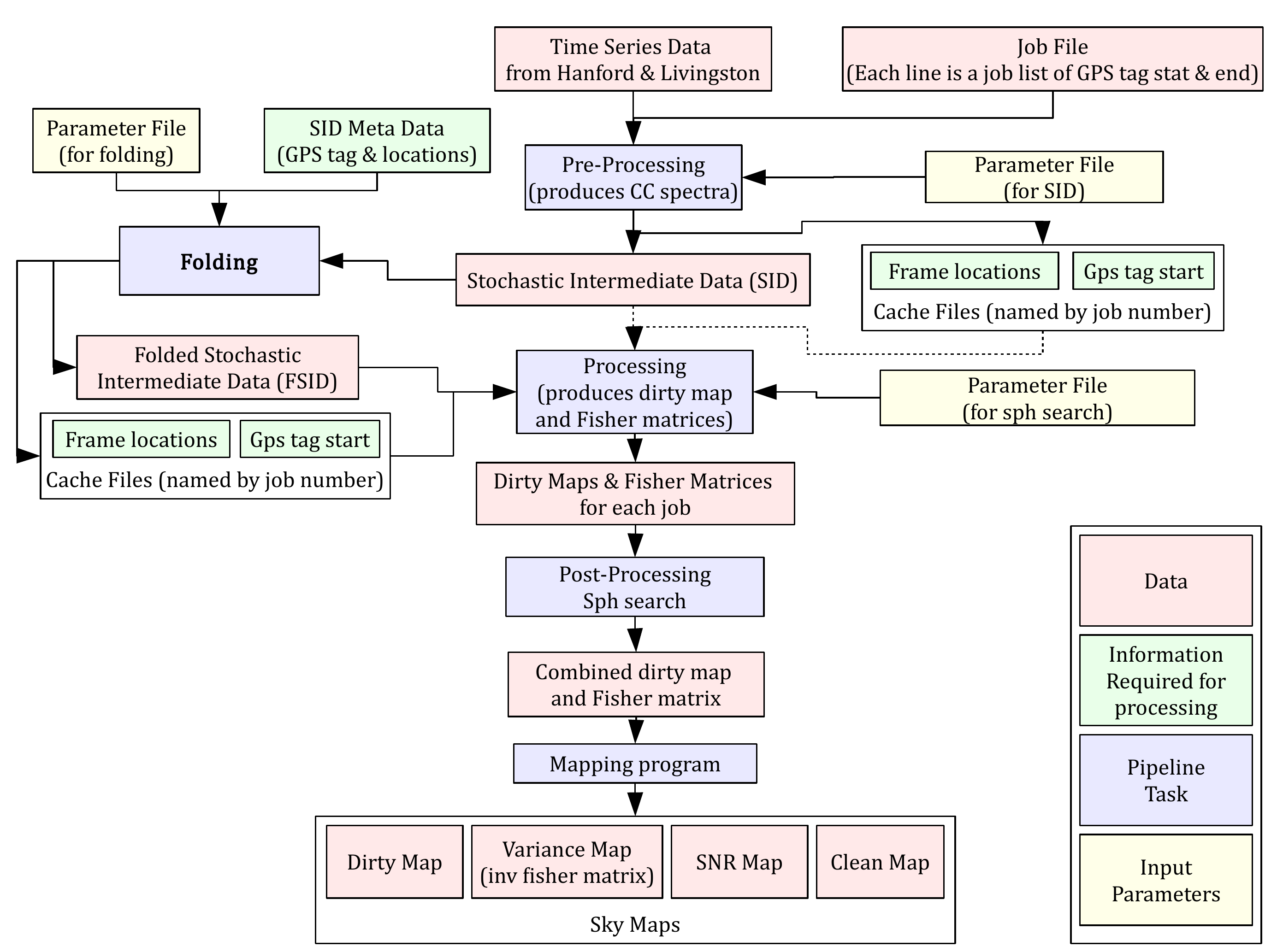}%
\caption{The flowchart illustrates the structure of the Stochastic pipeline and the path followed for the validation of results from unfolded and folded data. 
The main point to note here is that the same analysis was performed on SID and FSID frames and the FSID frames were produced from the same set of SID frames.}
\label{fig:flowchart}
\end{figure*}

Next, we produce FSID by folding the same set of SID frames. The SpH search is then applied to the FSID frames following the identical procedure as that for the SID frames and a skymap was generated. The only difference now is that we use far less computing power to process the data, as quantified below.
The procedure followed for this validation is depicted in the flow chart shown in Figure~\ref{fig:flowchart}.

We perform the validation for $\sim 10$ calendar days, $\sim 100$ calendar days and full $\sim 2$ calendar years worth of LIGO's fifth science run (S5) data from Hanford (H1) and Livingston (L1) detectors. Computation time required to fold the data was respectively $\sim 0.2, 2, 9$ CPU hours (depending on the load on the computer and storage devices at the time of the analysis) and the cost for performing the SpH analysis on the unfolded data are respectively $\sim 1, 10, 48$ CPU hours. Since folding is a one time affair, while radiometer analysis is run on the same data many times, computational cost for folding is practically negligible. The cost of running the SpH analysis on folded data is nearly {\em a constant}, it takes only $\sim 10$ minutes on a single core of a 2.0GHz Intel Xeon E7-4820 processor. Hence the computational cost saving is $\sim 300$ times for one full analysis of the S5 data (because S5 has total $\sim 300$ days of usable cross-power data) and this factor will remain the same for every analysis, isotropic or directed, that one performs on the same set of data, e.g., with different choices of $H(f)$, angular resolution, frequency bands and masks.

\begin{figure*}
\centering
\includegraphics[trim=0 80 0 60, clip=true, width=0.45\textwidth]{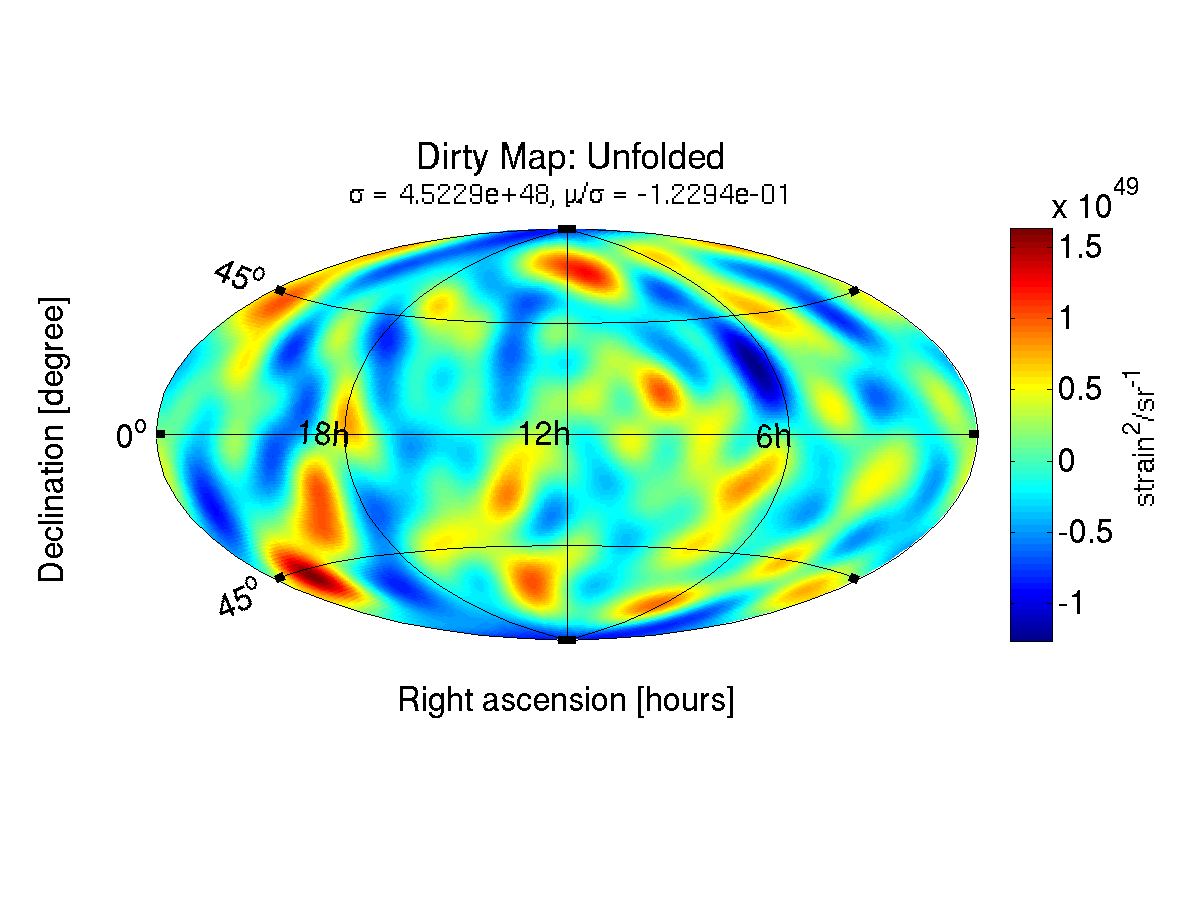}\qquad
\includegraphics[trim=0 80 0 60, clip=true,width=0.45\textwidth]{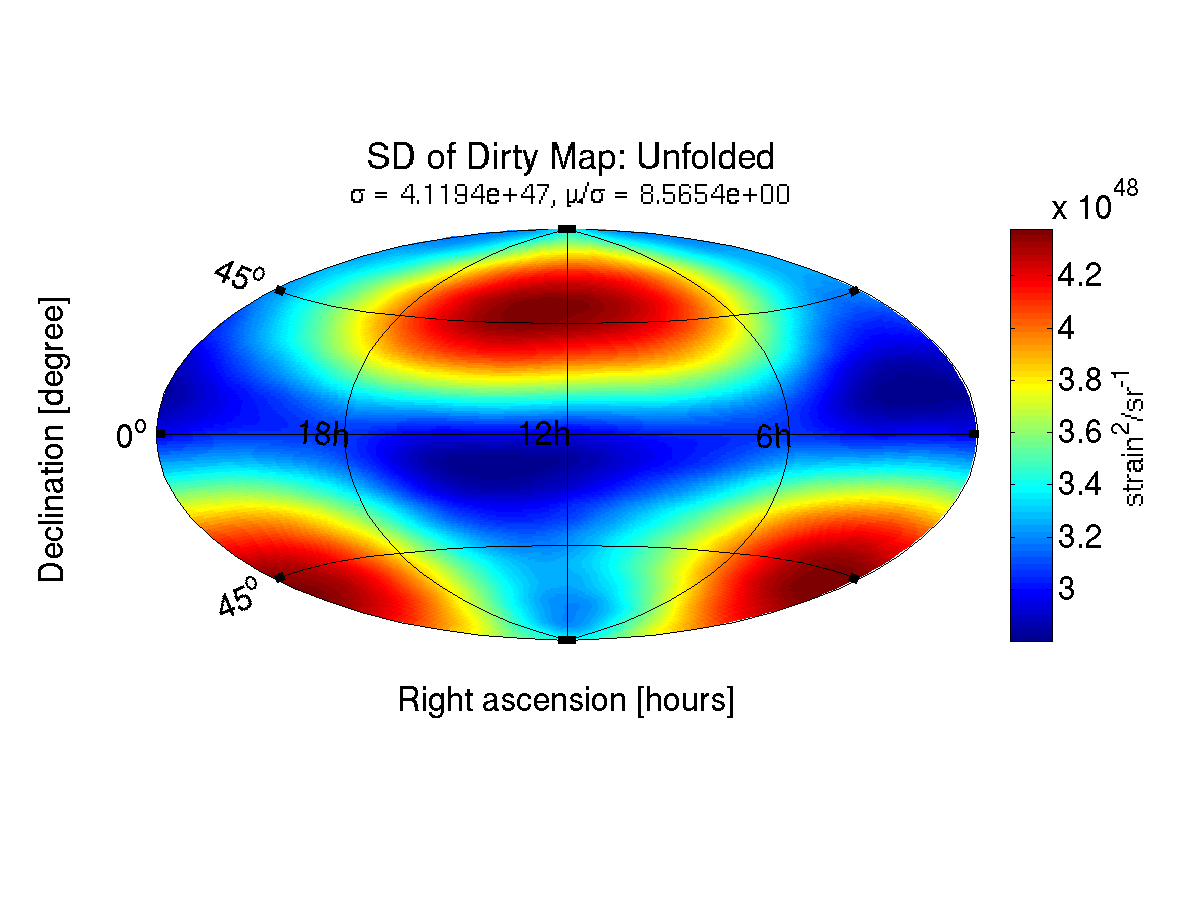}\\
\includegraphics[trim=0 80 0 60, clip=true,width=0.45\textwidth]{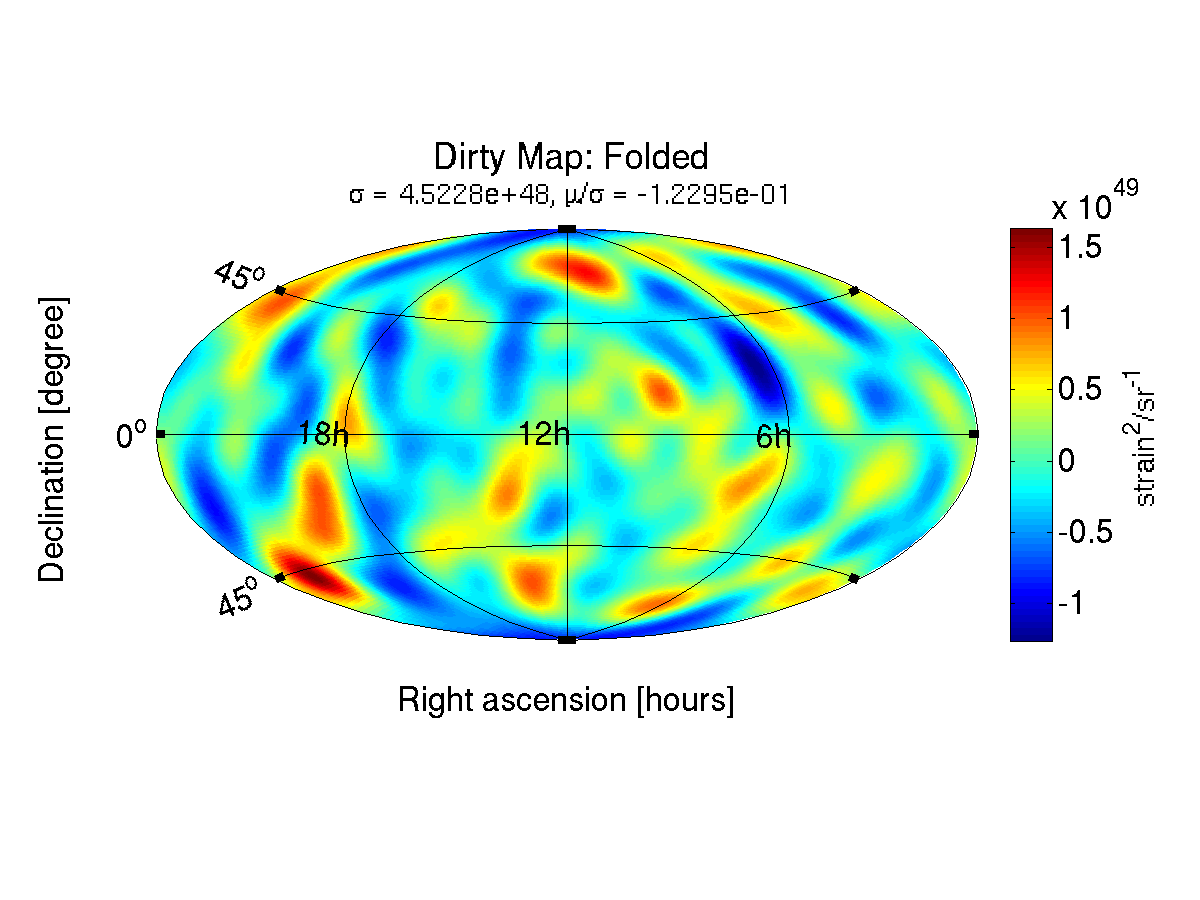}\qquad
\includegraphics[trim=0 80 0 60, clip=true,width=0.45\textwidth]{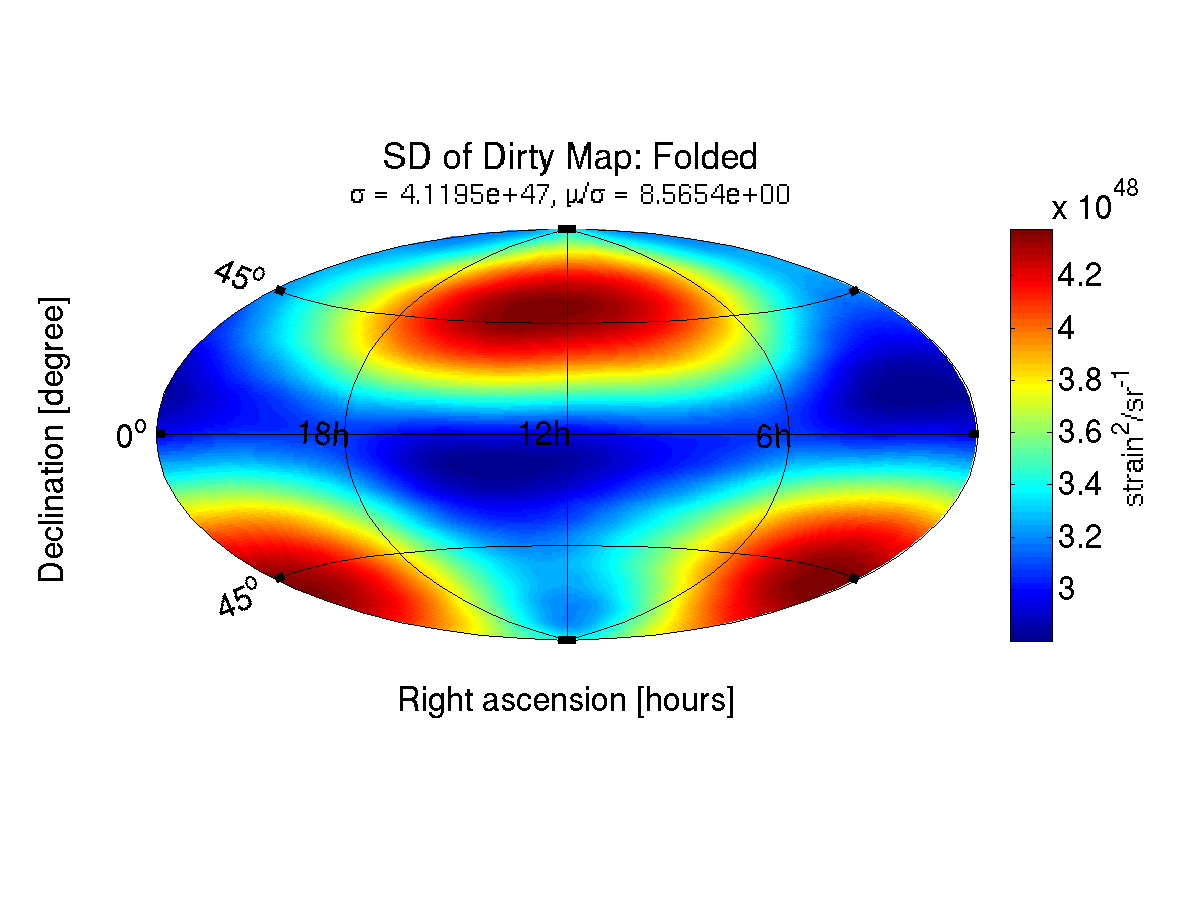}\\
\includegraphics[trim=0 80 0 60, clip=true,width=0.45\textwidth]{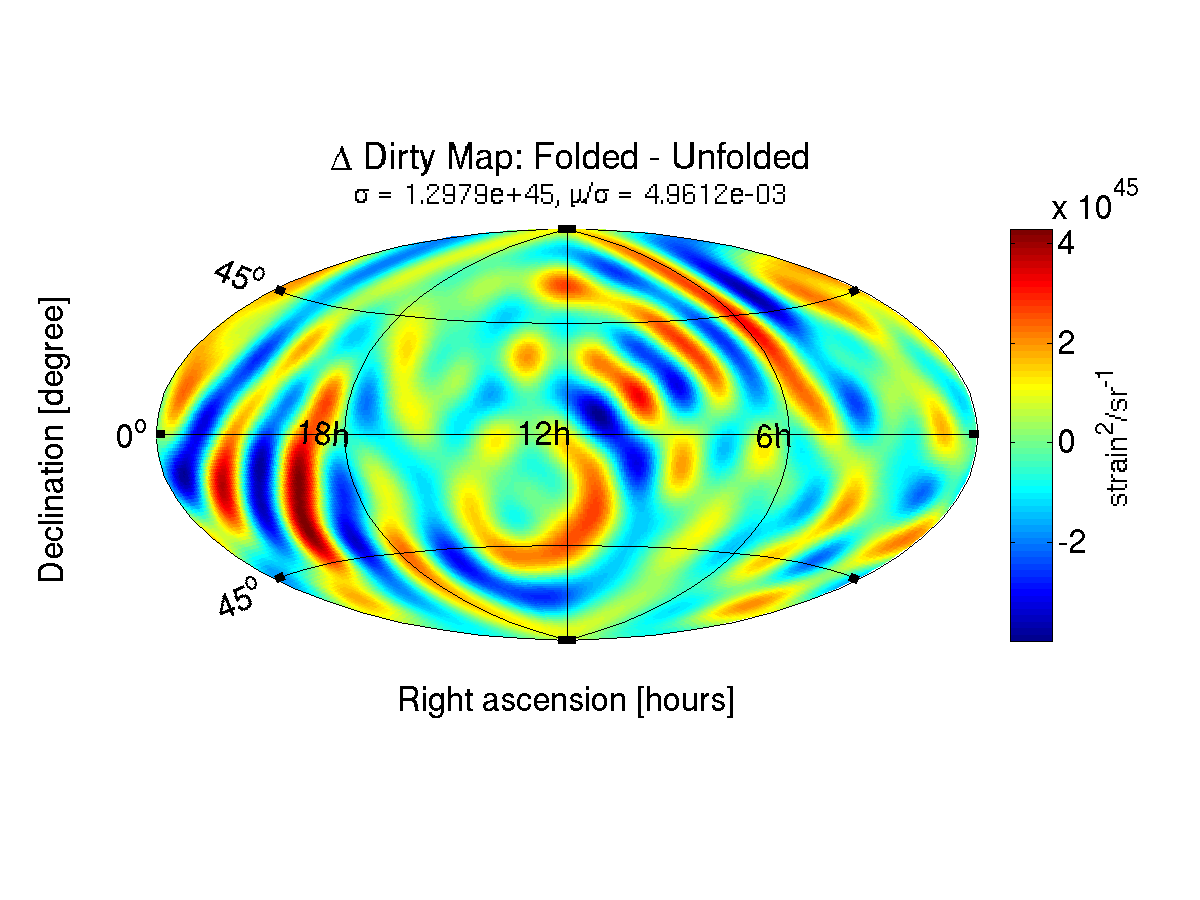}\qquad
\includegraphics[trim=0 80 0 60, clip=true,width=0.45\textwidth]{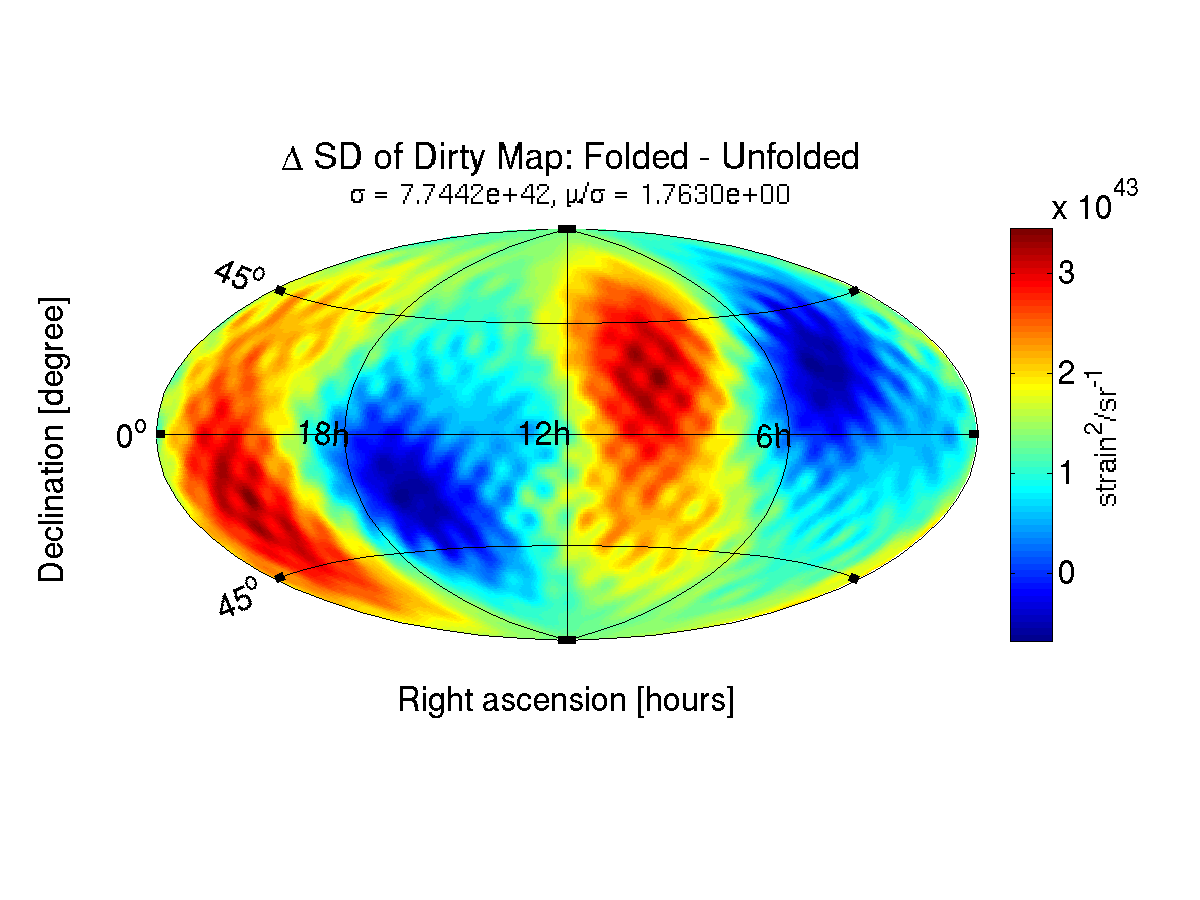}\\
\includegraphics[trim=0 80 0 60, clip=true,width=0.45\textwidth]{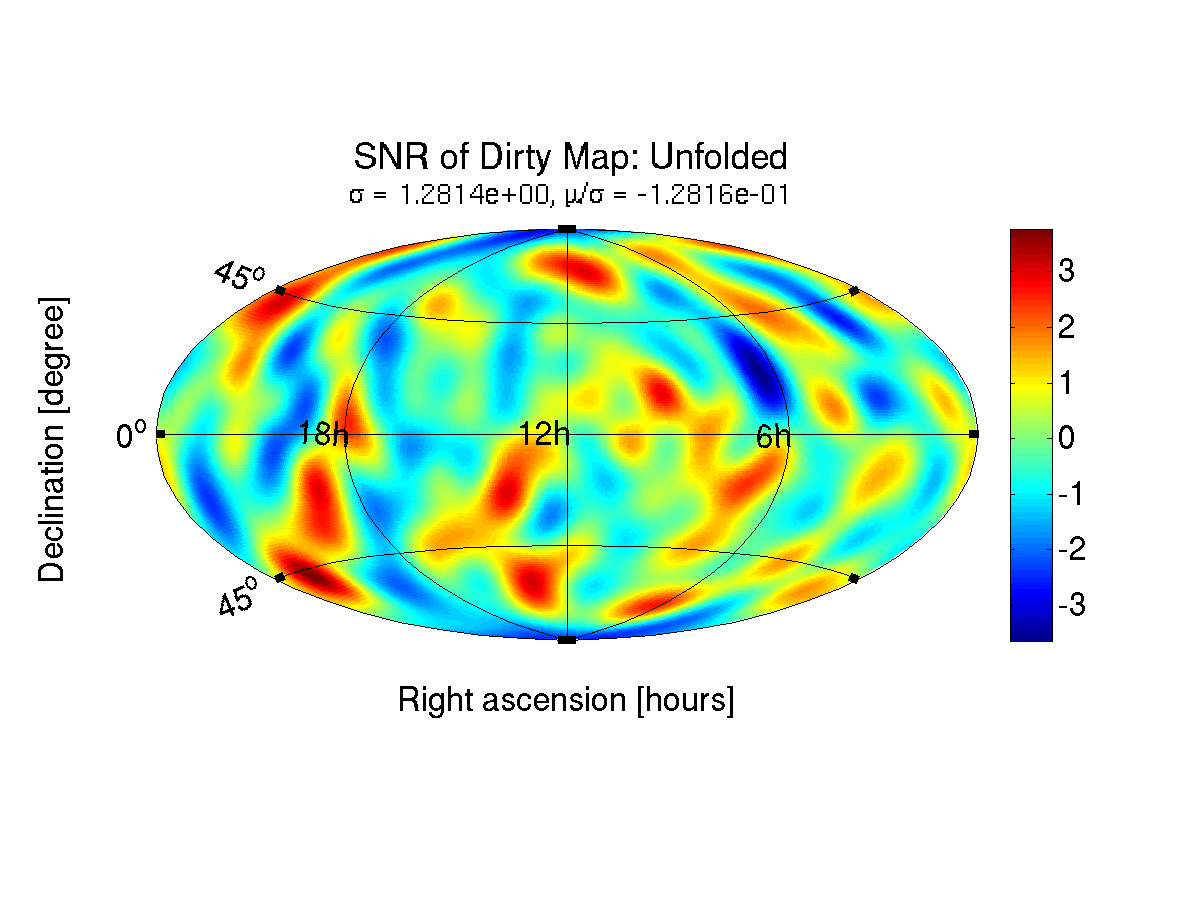}\qquad
\includegraphics[trim=0 80 0 60, clip=true,width=0.45\textwidth]{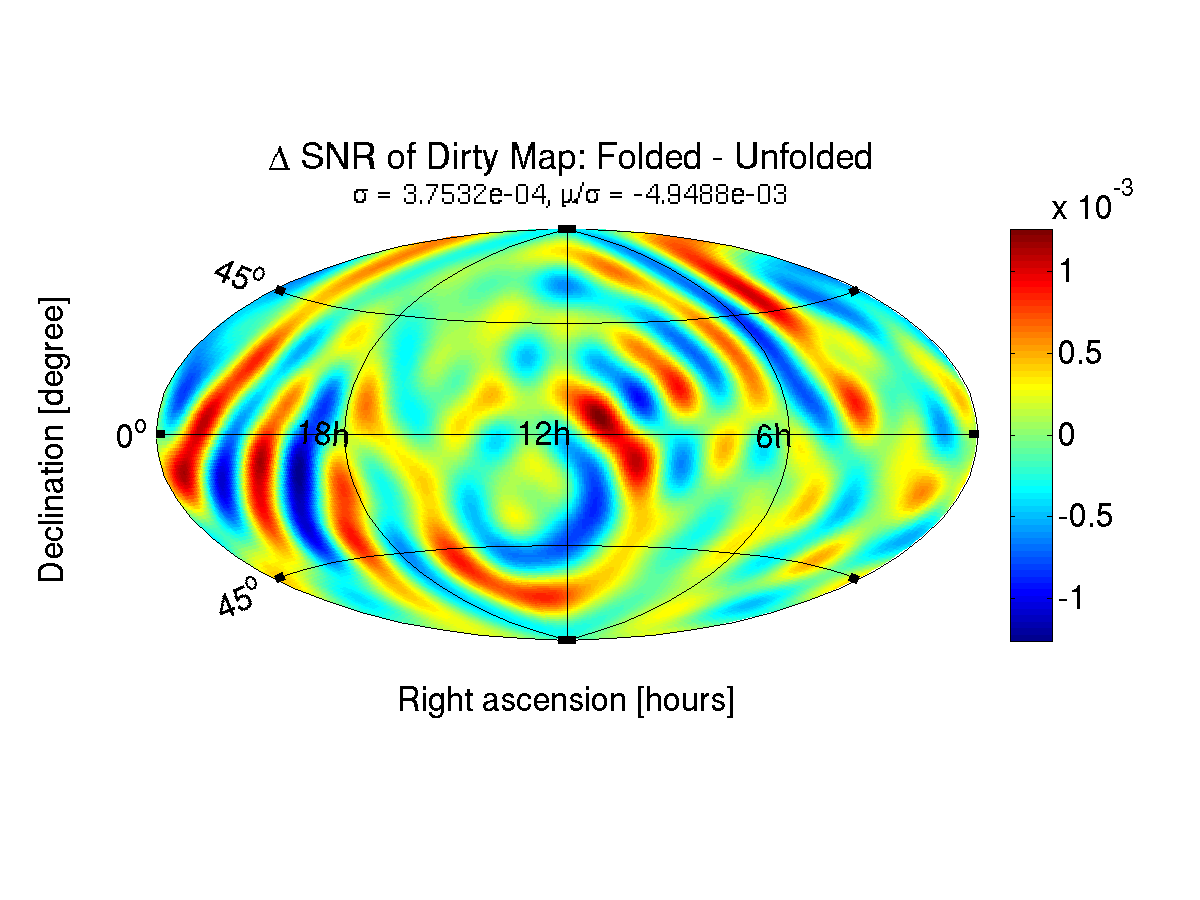}\\
\includegraphics[trim=0 80 0 60, clip=true,width=0.45\textwidth]{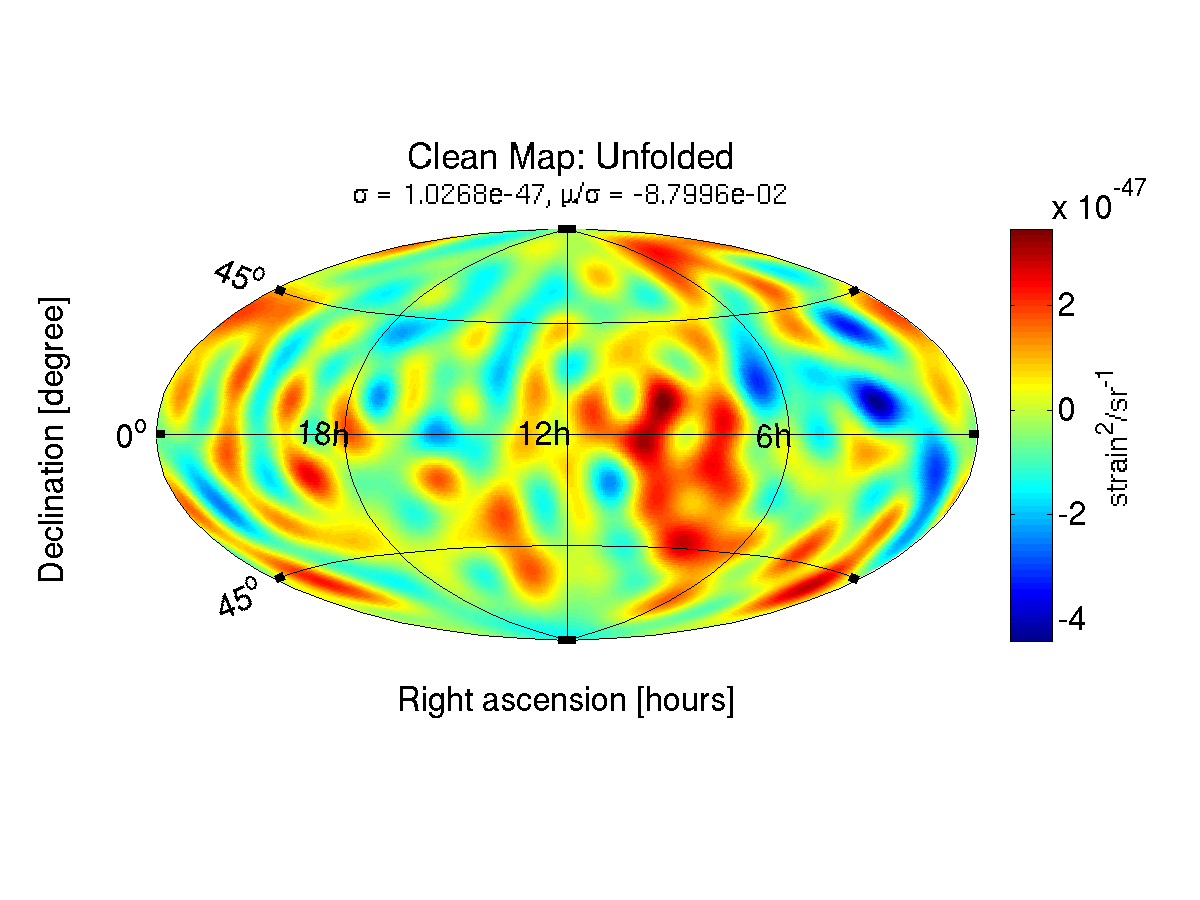}\qquad
\includegraphics[trim=0 80 0 60, clip=true,width=0.45\textwidth]{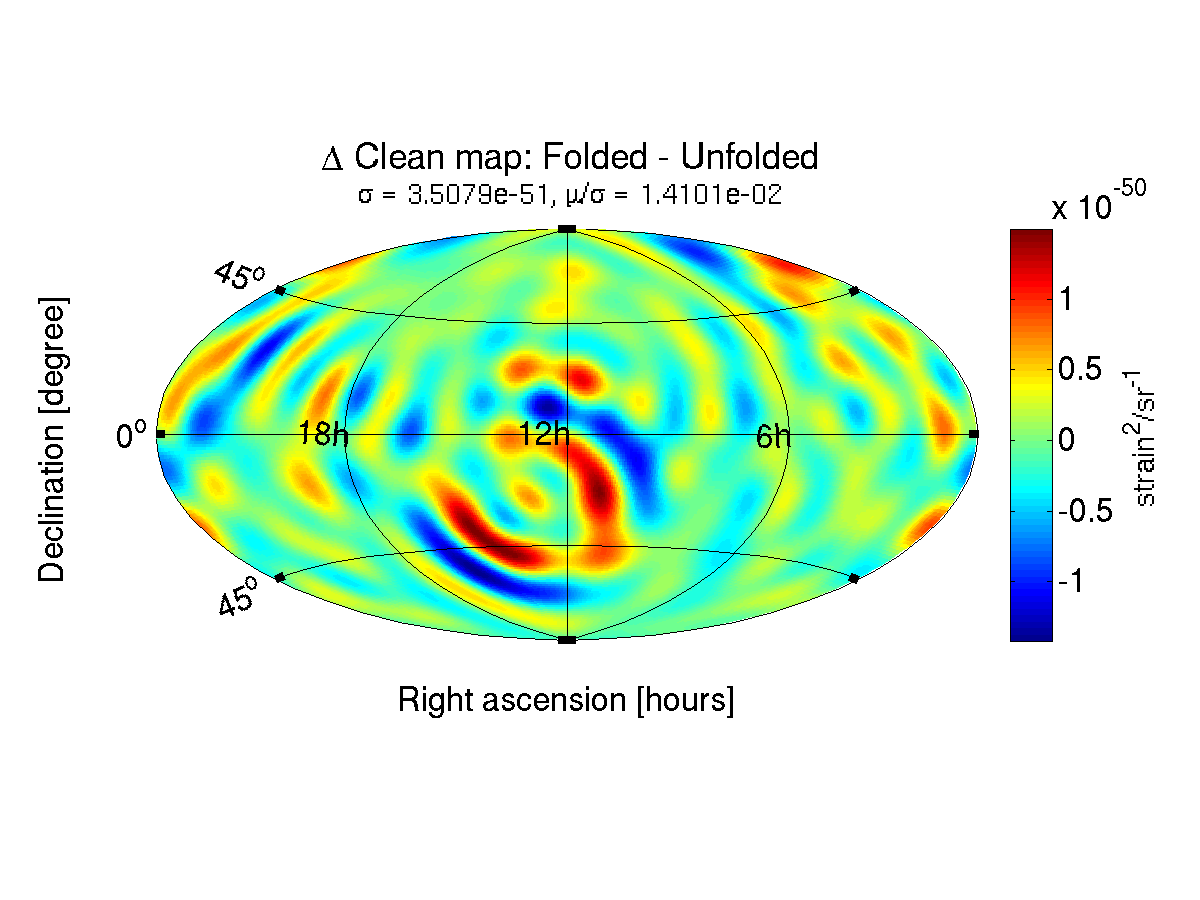}\\
\caption{\label{fig:Maps-not-folded}Maps constructed from 10 days of LIGO S5 unfolded and folded data and their differences are plotted here. Row \#1 and row \#2 show the dirty map (left) and its standard deviation map (right) obtained respectively from unfolded and folded data. The differences between the top two rows are plotted in row \#3. The results obtained from unfolded and folded data are clearly small. In row \#4 and \#5 we show the SNR and clean maps respectively from the unfolded data (left) and their differences (right) from the corresponding maps obtained from folded data. The differences are presented in a more quantitative form in Table~\ref{tab:diff}. Note that all the difference maps are statistically much smaller then the unfolded or folded maps, hence validating the folding method and the code.}
\label{fig:Map}
\end{figure*}

The comparison of results obtained from folded and unfolded data are presented in Figure~\ref{fig:Maps-not-folded} and Table~\ref{tab:diff}. Since the differences are larger for short durations of data, the figures are shown only for $\sim 10$ days' data. In Figure~\ref{fig:Maps-not-folded}, the dirty map (left) and its standard deviation map (right) for unfolded (row \#1) and folded (row \#2) data are shown. Row \#3 of Figure~\ref{fig:Maps-not-folded} shows the difference between the top two rows. Clearly, the maps from folded and unfolded data match very well and are visually indistinguishable. We also plot the SNR (row \#4) and deconvolved clean (row \#5)  maps from unfolded data (left) and their differences from the corresponding maps obtained from folded data (right). As can be seen from the colorbars, the differences are much smaller than the actual maps.

\begin{table*}
\begin{ruledtabular}
\begin{tabular}{lccc}
Observation time                                  & 10 calendar days   &  100 calendar days & Full S5\\
GPS start-end (sec)                               & 860832366-861701598 & 860832366-869499943 & 816065726-877591411 \\\hline
Real part of Fisher matrix          &  $2.55\times 10^{-5}$ & $1.02\times 10^{-5}$ & $5.57\times 10^{-6}$ \\	
Imaginary parts of Fisher matrix    &  $3.66\times 10^{-5}$ & $1.51\times 10^{-5}$ & $7.93\times 10^{-6}$ \\	
SpH coefficients of dirty map       &  $3.34\times 10^{-4}$ & $1.76\times 10^{-4}$ & $1.92\times 10^{-4}$ \\	
SpH coefficients of clean map       &  $3.44\times 10^{-4}$ & $2.74\times 10^{-4}$ & $2.62\times 10^{-4}$ \\	
Dirty map in pixel space            &  $2.85\times 10^{-4}$ & $1.53\times 10^{-4}$ & $1.67\times 10^{-4}$ \\	
Clean map in pixel space            &  $3.40\times 10^{-4}$ & $2.23\times 10^{-4}$ & $2.34\times 10^{-4}$ \\	
Standard deviation map of dirty map &  $4.42\times 10^{-6}$ & $2.80\times 10^{-6}$ & $1.82\times 10^{-6}$ \\	
SNR map of dirty map                &  $2.91\times 10^{-4}$ & $1.59\times 10^{-4}$ & $1.73\times 10^{-4}$ \\	
\end{tabular}
\end{ruledtabular}
\caption {\label{tab:diff} Table of fractional RMS differences between results for different quantities obtained from unfolded and folded data. If $\mathbf{A}$ and $\mathbf{B}$ are two result vectors (e.g., pixel space maps or Spherical Harmonic coefficients) obtained respectively from unfolded and folded data, the fractional RMS difference is given by $\| \mathbf{B} - \mathbf{A} \| / \|\mathbf{A} \|$, where $\|\mathbf{A} \| := \sqrt{\mathbf{A}^\dagger \cdot \mathbf{A}}$. The differences in the table are much smaller than $1$, implying excellent match between the results.  Some of the maps and differences for 10~days' data are shown in Figure~\ref{fig:Maps-not-folded}.}
\end{table*}

The quantitative differences are presented in Table-\ref{tab:diff}, which shows the fractional root-mean-square (RMS) differences between different maps for the three cases. If a result vector $\mathbf{A}$ is obtained from unfolded data and $\mathbf{B}$ is the corresponding result from folded data, we use the usual definition of fractional RMS difference, which is given by $\| \mathbf{B}-\mathbf{A}\| / \| \mathbf{A} \|$, where the norm of a $\mathbf{A}$ is defined as $ \| \mathbf{A} \| := \sqrt{\mathbf{A}^\dagger \cdot \mathbf{A}}$. Here each component of the vectors $\mathbf{A}$ and $\mathbf{B}$ corresponds to a pixel index or a pair of $l,m$ in the Spherical Harmonic basis. From Figure~\ref{fig:Maps-not-folded} and Table~\ref{tab:diff}, it is clear that the differences are much smaller than the standard deviation at each pixel (as evident from the SNR difference map) and, hence, can be ignored for all practical purposes.

It is worth mentioning that the match between the skymaps from unfolded and folded data that we see here is not just a statistical match, it is an absolute match. The whole skymaps obtained from the data sets match, not just their statistical properties like mean or variance. So the match implies that the signals have been combined with the right phases in the folded data. Therefore, performing simulations with injected signals would be a redundant exercise and hence not pursued here.

The residual difference of $\sim 10^{-4}$ in the dirty map is caused by misalignment between the SID and FSID frames. This difference is much smaller than the standard deviation, as evident from the SNR difference, and hence does not need any special attention in practice. However, since they arise from two avoidable reasons, we describe the cause and remedy for these misalignment for the sake of completeness. In our implementation we compute the Greenwich mean sidereal time for the mid-segment GPS timestamp of an SID frame and fold that data into an FSID frame which has the closest sidereal time. If all the SID frames were separated by multiples of $26$~sec, we could perfectly align every SID frame with its corresponding FSID frame, by choosing an appropriate offset (between $0-26$~sec) for the start time of the first FSID frame. However, the LIGO S5 SID frames are separated by multiples of $26$~sec {\em only} for contiguous stretches of data, spanning over a maximum of  few days. This does not hold for many days worth of data constituted by non-contiguous segments. Hence, in the later case, SID and FSID frames can not be perfectly aligned for every segment. One can, however, align the whole dataset easily by dropping on the average $\sim 5$ minutes of data per day. Secondly, a misalignment is also caused by the fact that one sidereal day is slightly longer than $86164$~sec by $\sim 0.1$~sec. This implies that there will be an alignment shift of $\sim 0.1$~sec per day,  which can accumulate up to $13$~sec then it resets to $-13$~sec in our implementation.  This error can be vastly reduced by choosing the segment duration more precisely, say, $52000054 \mu$s. Then the shift per sidereal day (given by the reminder when one mean sidereal day, $86164.090530833$~sec,  is divided by the segment duration) will be $1052.8 \mu$s, that is, less than a second for the whole duration of S5 run. Note that the baselines formed by the present and upcoming ground based laser interferometric detectors have an angular resolution of a few degrees ($\sim 0.1$ rad)~\cite{Mitra07}. In the worst case, an average timing jitter of $\delta T  = 13$~sec corresponds to an angular shift of $\sim 2 \pi \delta T / \sday \approx 0.001$, which is $\sim 1$\% of the angular resolution of the radiometers considered here and hence these timing inaccuracies are practically negligible.

In summary, the results demonstrate that the folding method and the codes are working correctly. The main difference was that to analyse full S5 data, we processed nearly a million SID frames, while only 3314 FSID frames were analysed to get the same result. This reduced the computation cost for processing and post-processing by a factor of $\sim 300$.


\section{Conclusions}
\label{concl}

We have formulated and implemented an algorithm to fold entire datasets for pairs of GW detectors to only one sidereal day's data, which is of enormous advantage to GW radiometer analyses. We developed a parallel pipeline to implement this method on data from ground-based interferometers and applied to LIGO's fifth science run data. The skymaps obtained from folded and unfolded data through a standard LIGO anisotropic SGWB search pipeline showed excellent match, thus convincingly validating the method and the implementation.

Folding follows from an algebraic identity, hence the results are exact. Even for incorporating overlapping windows, the tiny error arising from the inversion of a tridiagonal matrix is inherent to the analysis, folded data merely incorporates that approximation with matching precision. Also, folding process takes care of the correction, so that, folded data can be analysed in a straightforward way without repeating this complex procedure, as if there are no overlapping windows. 

The folding process, though fairly straightforward in principle, does capture the complications involved in real data analysis. For instance,  it {\em does not assume stationarity} of noise, hence no special care is necessary to account for variation in noise power spectrum, say, on different days of the week or at different times of the day. Folded data also account for the overlapping Hanning windows applied to data to reduce spectral leakage. The frequency masks, often used in analyses to remove the noisy bits of data, can be readily applied to folded data. One can also remove non-stationary data~\cite{sgwbS4dir} while folding, by discarding those frequency bins of an unfolded segments, whose variances are significantly (say, more than $ 20$\%) different from those in the adjacent segments.

The advantages one can derive out of the folded data is many-fold:
\begin{enumerate}
\item Efficiency: If $n$ sidereal days' data is folded to one sidereal day, computational cost reduces by a factor of $n$. The one time folding step requires negligible amount of computing compared to the total  cost of running different radiometer analysis few times each on the same data. Hence for few years worth of data, {\em the computational cost reduction could be by a factor of as much as $\sim 1000$}. Moreover, the folding step already accounts for overlapping window correction and other preprocessing steps, so there is further reduction in computation cost.
%
%
\item Portability: Since folded data volume is reduced by a factor of $n$, it will be convenient to transport data from computer to computer. For standard stochastic analysis, folded data size is only $\sim 1.3$ gigabytes, which comfortably fits on a USB memory stick.
\item Convenience: The above two points imply that folded data for standard searches can be processed in a personal computer. This will allow multiple experimentation with current analyses and will make it easier to develop new search algorithms. In addition, cross-correlation based searches which were not conceivable due to computational limitations have now become possible to perform.
\item Robustness: Since the preprocessing steps are the same for all analyses with folded data, better consistency check between different searches would be possible.
\item Modularity: Since the folding part is separate from the search part, it will be possible to do the disk I/O intensive folding part in a low level language (e.~g., {\tt C}) and the complicated algebra of filtering for different searches in more high level language like {\tt MATLAB} or {\tt Python}.
\item Management: Folded data duration and volume are independent of the total observation time, hence the computation cost, storage etc. are independent of the number of days of observation. Which can be of help in planning computation budget and designing of parallel pipelines.
\end{enumerate}

%
%
%

The enormous efficiency that folded data brings will allow one to perform different kinds of neat analyses. For instance, an all sky targeted search for narrowband sources over a frequency range of few hundred Hz may become feasible. One may be able to push the frequency bin size down in order to efficiently search for narrow line emissions~\cite{noCoarseGrain}. We caution the reader that we do not fold single detector's data, which could perhaps provide enormous efficiency to search for purely monochromatic (coherent) signals. The folding scheme described here applies exclusively to analyses where data from two different detectors are cross-correlated, which can sometimes be used to optimally or sub-optimally search for narrowband sources. On the other hand, for long-duration broadband sources, folding may be coupled with coarse-graining~\cite{coarseGrain} for even more efficiency. 
In conclusion, GW radiometer is the optimal analysis for detecting long duration unknown sources. Folded data will enable enough experimentation with the radiometer analysis  to look for such sources.

\begin{acknowledgments}
We would like to thank Albert Lazzarini for important comments, suggestions and encouragement. This work also benefitted from discussions with Tania Regimbau, Joe Romano, John Whelan, Dipongkar Talukder, Eric Thrane, Nelson Christensen, Patrick Meyers, Tarun Souradeep, Sanjeev Dhurandhar, Sukanta Bose and the stochastic group of the LIGO Virgo Scientific Collaboration for useful discussions. AA acknowledges the support of Council of Scientific and Industrial Research (CSIR), India. SM acknowledges the support of Science and Engineering Research Board (SERB), India for the FastTrack grant SR/FTP/PS-030/2012.
This document has a LIGO preprint number LIGO-P1500007.
\end{acknowledgments}

\bibliography{folding}

\end{document}